\documentclass{article}
\usepackage{emulateapj,apjfonts,rotating,epsf}

\slugcomment{(Accepted for publication in ApJ)}
\lefthead{Mass Spectra from Turbulent Fragmentation }
\righthead{Ralf\ S.\ Klessen}

\begin{document}

\title{The Formation of Stellar Clusters: Mass Spectra from Turbulent
 Molecular Cloud Fragmentation}
 \author{Ralf S.\ Klessen}
\affil{UCO/Lick Observatory, University of California at
 Santa Cruz, Santa
 Cruz, CA 95064, USA ~~~(e-mail: ralf@ucolick.org)\\Max-Planck-Institut f{\"u}r
 Astronomie, K{\"o}nigstuhl 17, 69117 Heidelberg, Germany}

\begin{abstract}
  Star formation is intimately linked to the dynamical evolution of
  molecular clouds. Turbulent fragmentation determines where and when
  protostellar cores form, and how they contract and grow in mass via
  competitive accretion from the surrounding cloud material.  This
  process is investigated, using numerical models of self-gravitating
  molecular cloud dynamics, where no turbulent support is included,
  where turbulence is allowed to decay freely, and where it is
  continuously replenished on large, intermediate and small scales,
  respectively. Molecular cloud regions without turbulent driving
  sources, or where turbulence is driven on large scales, exhibit
  rapid and efficient star formation in a clustered mode, whereas
  interstellar turbulence that carries most energy on small scales
  results in isolated star formation with low efficiency.

  The clump mass spectrum of shock-generated density fluctuations in
  pure hydrodynamic, supersonic turbulence is not well fit by a power
  law, and it is too steep at the high-mass end to be in agreement
  with the observational data. When gravity is included in the
  turbulence models, local collapse occurs, and the spectrum extends
  towards larger masses as clumps merge together, a power-law
  description $dN/dM \propto M^{\nu}$ becomes possible with slope $\nu
  \lesssim -2$. In the case of pure gravitational contraction, i.e.\
  in regions without turbulent support, the clump mass spectrum is
  shallower with $\nu \approx -3/2$. 

  The mass spectrum of protostellar cores in regions without turbulent
  support and where turbulence is replenished on large-scales,
  however, is well described by a log-normal or by multiple power
  laws, similar to the stellar IMF at low and intermediate masses. The
  model clusters are not massive enough to allow for comparison with
  the high-mass part of the IMF. In the case of small-scale
  turbulence, the core mass spectrum is too flat compared to the IMF
  for all masses.
\end{abstract}


\section{Introduction}
Understanding the processes that lead to the formation of stars is one
of the fundamental challenges in astronomy.  As mass is the dominant
parameter determining stellar evolution, reproducing and explaining
the initial mass function of stars (IMF) is a key requisite for any
realistic  theory of star formation.

Stars are born in turbulent interstellar clouds of molecular hydrogen.
The location and the mass growth of young stars are hereby intimately
coupled to the dynamical cloud environment. Stars form by
gravitational collapse of shock-compressed density fluctuations
generated from the supersonic turbulence ubiquitously observed in
molecular clouds (e.g.\ Elmegreen 1993, Padoan 1995, Klessen, Heitsch,
\& Mac~Low 2000, Padoan et al.\ 2001). Once a gas clump becomes
gravitationally unstable, it begins to collapse and the central
density increases considerably, giving birth to a protostar. In this
dynamic picture, star formation takes place roughly on a free-fall
timescale, as opposed to the ``standard'' model of the inside-out
collapse of singular isothermal spheres, where core formation is
dominated by the ambipolar diffusion timescale (Shu 1977, Shu, Adams,
\& Lizano 1987).  Altogether, star formation can be seen as a
two-phase process: First, {\em turbulent fragmentation} leads to
transient clumpy molecular cloud structure, with some of the density
fluctuation exceeding the critical mass and density for gravitational
contraction.  Second, the {\em collapse of individual Jeans-unstable
protostellar clumps} builds up the stars.  In this phase, a nascent
protostar grows in mass via accretion from the infalling envelope
until the available gas reservoir is exhausted or stellar feedback
effects become important and remove the parental cocoon --- a new star
is born (e.g.\ Andr{\'e}, Ward-Thompson, \& Barsony 2000, Myers,
Evans, \& Ohashi 2000).  The terms shock-generated density
fluctuations and gas clumps are used synonymously, and clumps are
identified using a 3-dimensional clump-finding algorithm comparable to
the one described in Williams, De~Geus, \& Blitz (1994). Protostellar
cores in the simulations are defined as the (unresolved) high-density
central regions of collapsing clumps, where individual protostars
build up.

\begin{deluxetable}{crccccl}
\tablehead{
\colhead{model} &
\colhead{type}  &
\colhead{$k$\tablenotemark{a}} &
\colhead{${\cal M}_{\rm rms}$\tablenotemark{b}} &
\colhead{$N_{\rm J}$\tablenotemark{c}} &
\colhead{particles} &
\colhead{further reference\tablenotemark{d}}
}
\tablecaption{\label{tab:models}
Properties of the Considered Molecular Cloud Models}
\startdata
1 & Gaussian density& --- & --- & 220 & $5\times 10^5$ & model $\cal I$ in
KB \\
2 & decaying turbulence & $[1\dots8]$ & [5.0] & 220 & $2\times 10^5$ & --- \\
3 & driven turbulence & $1\dots2$ & 3.3 & 120 & $2\times 10^5$ & model
${\cal A}1h$ in KHM \\
4 & driven turbulence & $3\dots4$ & 3.3  & 120 & $2\times 10^5$ & model ${\cal A}2h$ in
KHM \\
5 & driven turbulence & $7\dots8$ & 3.3 & 120 & $2\times 10^5$ & model ${\cal A}3h$ in
KHM \\
\enddata
\tablenotetext{a}{Driving wavenumber (in model 2, driving stopped 
when gravity is `turned on' at the stage of fully established turbulence) }
\tablenotetext{b}{Mach number in turbulent equilibrium as calculated from the
1-dimensional rms
velocity dispersion, ${\cal M}_{\rm rms} \equiv \sigma_{\rm
1D}/c_{\rm s}$, where $c_{\rm s}$ is the sound speed and $\sigma_{\rm
1D}$ is defined via $\sigma_{\rm 1D}^2 = 1/3\,\sigma_{\rm 3D}^2 =
1/3\;(E_{\rm kin}/2)$ with $E_{\rm kin}$ being the total kinetic
energy. Recall that the total mass is unity. For  model 2 the value
corresponds to time $t<0$ before the driving mechanism was `turned
off' and  turbulence was allowed to decay.}
\tablenotetext{c}{Number of (spherical) mean thermal Jeans masses contained
in the system; note that this number is lower by 2 when using a cubic
definition of the Jeans mass}
\tablenotetext{d}{Corresponding model name in KB (paper I) and in KHM
(Klessen et al.\ 2000) for further details} 
\end{deluxetable}

Stars form in small aggregates or larger clusters (Lada 1992, Mizuno
et al.\ 1995, Testi, Palla, \& Natta 1998, also Adams \& Myers 2001),
where the interaction of protostellar cores and their competition for
mass from their surrounding may become important for shaping the
distribution of the final star properties (Bonnell et al.\ 1997,
Klessen, Burkert, \& Bate 1998, and Klessen \& Burkert 2000, 2001, in the
following papers I and II, respectively). This complex evolutionary
sequence involves a wide variety of different physical phenomena, and
it is not at all well understood which processes dominate and
determine the stellar mass spectrum.

The current investigation is the fourth in a series which focuses on the
first phase of the star formation process, modeling the turbulent
fragmentation of large subvolumes inside molecular clouds and the
dynamical evolution towards the formation of clusters of protostellar
cores.  Numerical simulations (\S2) of self-gravitating isothermal
gas, without turbulent support, with decaying turbulence, and with
supersonic compressible turbulence that is driven on large,
intermediate, and small scales, are used to analyze the star formation
resulting from the interplay between gravity on the one side and gas
pressure and turbulent motions on the other (\S3). In particular, the
relation between the masses of molecular clumps, protostellar cores
and the final stars in the considered models are discussed (\S4) and
the results summarized (\S5).

\setcounter{footnote}{0}
\section{Numerical Method and Driven Turbulence}
\label{sec:numerics}
To adequately describe turbulent fragmentation and the formation of
protostellar cores, it is necessary to resolve the collapse of
shock-compressed regions over several orders of magnitude in
density. Due to the stochastic nature of supersonic turbulence, it is
not known in advance where and when local collapse occurs. Hence, SPH
({\em smoothed particle hydrodynamics}) is used to solve the equations
of hydrodynamics. It is a Lagrangian method, where the fluid is
represented by an ensemble of particles and flow quantities are
obtained by averaging over an appropriate subset of the SPH particles
(Benz 1990). The method is able to resolve large density contrasts as
particles are free to move and so naturally the particle concentration
increases in high-density regions.  SPH can also be combined with the
special-purpose hardware device GRAPE (Sugimoto et al.\ 1990,
Ebisuzaki et al.\ 1993; also Steinmetz 1996) permitting calculations
at supercomputer level on a normal workstation. The simulations
presented here concentrate on subregions within a much larger cloud,
therefore periodic boundary conditions are adopted (Klessen
1997). Once the high-density, protostellar cores in the centers of
collapsing gas clumps exceed a density limit four orders of magnitude
above the mean density, they are substituted by `sink' particles
(Bate, Bonnell, \& Price 1995). These particles have the ability to
accrete gas from their envelopes, while keeping track of mass and
linear and angular momentum. By adequately replacing high-density core
with sink particles one is able to follow the dynamical evolution of
the system over many free-fall times.
 
The large observed linewidths in molecular clouds imply the presence
of supersonic velocity fields that carry enough energy to
counterbalance gravity on global scales (Williams, Blitz, \& McKee
2000).  However, it is known that turbulent energy dissipates rapidly,
i.e.\ roughly on the free-fall timescale (Mac Low et al.\ 1998, Stone,
Ostriker, \& Gammie 1998, Padoan \& Nordlund 1999).  To prevent or
considerably postpone global collapse, turbulence is required to be
continuously replenished.  This is achieved here by applying a
non-local driving scheme, that inserts energy in a limited range of
wavenumbers such that the total kinetic energy contained in the system
remains constant and compensates the gravitational contraction on
global scales (Mac~Low 1999).  The models do not include magnetic
fields, as their presence cannot halt the decay of turbulence (Mac~Low
et al.~1998, Stone et al.\ 1998, Padoan \& Nordlund 1999) and does not
significantly alter the efficiency of local collapse for driven
turbulence (Heitsch, Mac Low, \& Klessen 2001). Furthermore, possible
feedback effects from the star formation process itself (like bipolar
outflows, stellar winds, or ionizing radiation from new-born O or B
stars) are neglected.  This necessarily limits the interpretation of
the mass spectra at very late evolutionary stages of the system.
Hence, the current analysis restrains itself to phases when the mass
accumulated  protostellar cores is less than $\sim70$\% of the total
mass in the considered volume.

Altogether, five different models of molecular cloud dynamics are
considered here: To compare with the case of driven turbulence, model
1 describes the dynamical evolution of an initially Gaussian density
fluctuation field where turbulence is assumed to have already decayed
in the considered molecular cloud region. This describes the most
extreme case of clustered star formation, and the simulation is
identical to model $\cal I$ in paper I. The power spectrum of the
initial density fluctuations is $P(k)\propto k^{-2}$. Model 2 starts
with a fully established supersonically turbulent velocity field, but
turbulence is allowed to decay freely. Driven turbulence is
represented by model 3 where the energy source acts at wavenumbers $k$
in the range $1 \le k \le 2$, by model 4 which has $3 \le k \le 4$,
and model 5 with $7 \le k \le 8$. The wavelengths of the corresponding
perturbations are $\ell = L/k$, where $L$ is the total size of the
computed volume. Hence, kinetic energy is continuously added on large,
intermediate, and small scales, respectively. The driving strength is
adjusted to yield the same constant turbulent Mach number ${\cal
M}_{\rm rms} = 3.3$ for all three models (see Klessen et al.\ 2000,
models ${\cal A}1h$, ${\cal A}2h$, and ${\cal A}3h$).  Turbulence that
is driven on large scales appears to yield most appropriate
description of molecular cloud dynamics and star formation as is
suggested by statistical analysis of molecular cloud structure (e.g.\
Ossenkopf \& Mac Low 2001). The relevant model parameters are listed
in table \ref{tab:models}.

The models presented here are computed in normalized units. If scaled
to mean densities of $n({\rm H}_2) = 10^5\,$cm$^{-3}$, a value typical
for star-forming molecular cloud regions (e.g.\ in $\rho$-Ophiuchus,
see Motte, Andr{\'e}, \& Neri 1998) and a temperature of 11.4$\,$K
(i.e.\ a sound speed $c_{\rm s} = 0.2\,$km$\,$s$^{-1}$), the total
mass contained in the computed volume in models 1 and 2 is
220$\,$M$_{\odot}$ and the size of the cube is $0.34\,$pc. This
corresponds to 220 thermal Jeans masses. Models 3 to 5 have a mass of
120$\,$M$_{\odot}$ within a volume of ($0.29\,{\rm pc})^3$, equivalent
to 120 thermal Jeans masses\footnote{Throughout this paper the
  spherical definition of the Jeans mass is used, $M_{\rm J} \equiv
  4/3\,\pi \rho \lambda_{\rm J}^3$, with density $\rho$ and Jeans
  length $\lambda_{\rm J}\equiv \left(\frac{\pi{\cal R}T }{G
      \rho}\right)^{1/2}$, where $G$ and $\cal R$ are the
  gravitational and the gas constant. The mean Jeans mass $\langle
  M_{\rm J} \rangle$ is then determined from average density in the
  system $\langle \rho \rangle$. An alternative cubic definition,
  $M_{\rm J} \equiv \rho (2\lambda_{\rm J})^3$, would yield a value
  roughly twice as large.}. In the adopted scaling, the mean thermal
Jeans mass in all models is thus $\langle M_{\rm J} \rangle =
1\,$M$_{\odot}$, the global free-fall timescale is $\tau_{\rm ff} =
10^5\,$yr, and the simulations cover a density range from
$n({\rm H}_2) \approx 100\,$cm$^{-3}$ in the lowest density regions to
$n({\rm H}_2) \approx 10^9\,$cm$^{-3}$ where collapsing protostellar
cores are identified and converted into `sink' particles in the code.
In this density regime gas cools very efficiently and it is possible
to use an effective polytropic equation-of-state in the simulations
instead of solving the detailed radiation transfer equations. The
effective polytropic index is typically close to unity, $\gamma_{\rm
  eff} \lesssim 1$, except for densities $10^5\,$cm$^{-3} < n({\rm
  H}_2) < 10^7\,$cm$^{-3}$, where smaller values of $\gamma_{\rm eff}$
are expected (Spaans \& Silk 2000). For simplicity, a value of
$\gamma_{\rm eff}=1$, i.e.\ an isothermal equation of state, is
adopted for all densities in the simulations. Concerning the gas
temperature, this approximation is certainly valid as in star forming
clouds the temperature cannot drop significantly below the adopted
canonical value of 10$\,$K, even in the regime $10^5\,$cm$^{-3} <
n({\rm H}_2) < 10^7\,$cm$^{-3}$. However, it needs to be noted that
the stiffness of the equation of state also determines the density contrast
in shock compressed gas, and hence influences  the overall density
distribution in supersonic flows.  For further discussions see Scalo et al.\ (1998),
Ballesteros-Paredes, V{\'a}zquez-Semadeni, \& Scalo (1999), and Spaans
\& Silk (2000). Variations in $\gamma_{\rm eff}$ influence  the local Jeans
scale in shock-compressed density fluctuations and may modify the resulting
mass spectrum of collapsing cores. This effect needs to be investigated in
more detail.

\begin{figure*}[p]
\unitlength1cm
\begin{picture}(18.0,22.9)
\put( 0.0, -1.5){\epsfxsize=17cm \epsfbox{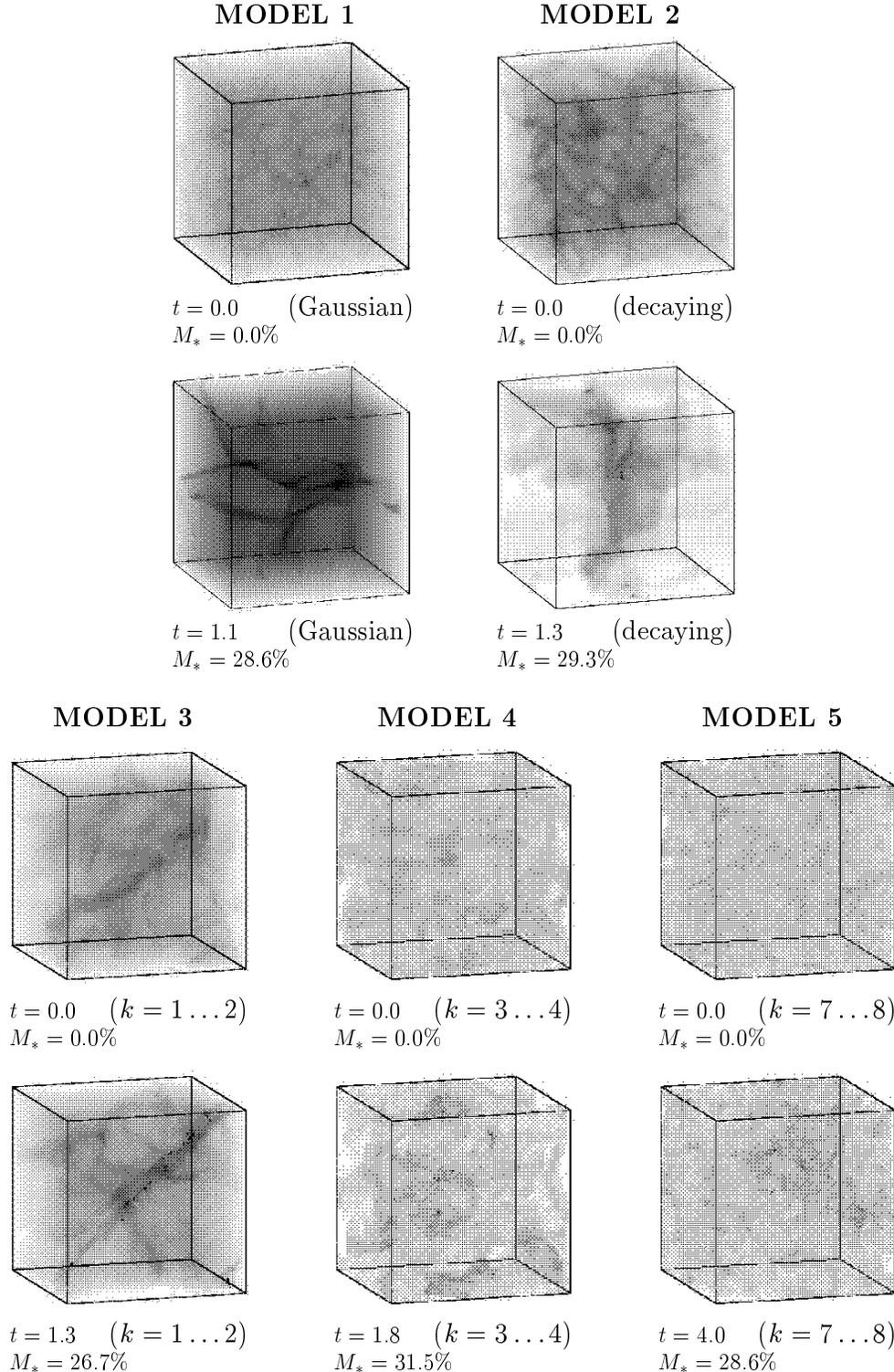}}
\end{picture}
\caption{\label{fig:3Dplot} Comparison of the
  gas distribution in the five models at two different phases of the
  dynamical evolution, at $t=0$ indicating the initial density
  structure, just before gravity is `switched on', and after the first
  cores have formed and accumulated roughly $M_{\rm *} \approx 30$\%
  of the total mass. The high-density (protostellar) cores are
  indicated by black dots. Note the different time interval needed to
  reach the same dynamical stage. Time is normalized to the global
  free-fall timescale of the system, which is $\tau_{\rm ff} =
  10^5\,$yr for $T=11.4\,$K and $n({\rm H}_2) = 10^5\,$cm$^{-3}$. The
  cubes contain masses of $220\,\langle M_{\rm J}\rangle$ (models 1
  and 2) and $120\,\langle M_{\rm J}\rangle$
  (models 3 to 5), respectively, where the average thermal Jeans mass
  is $\langle M_{\rm J}\rangle= 1\,$M$_{\odot}$ with the above scaling. The
  considered volumes are $(0.32\,$pc$)^3$ and $(0.29\,$pc$)^3$,
  respectively. Note, however, that the isothermal models are freely scalable as
  discussed in \S\ref{sec:numerics}.}
\end{figure*}

\section{Star Formation from Turbulent Fragmentation}
\label{sec:location-time}

Stars form from turbulent fragmentation of molecular cloud material.
Supersonic turbulence, even if it strong enough to counterbalance
gravity on global scales, will usually {\em provoke} local collapse.
Turbulence establishes a complex network of interacting shocks, where
converging shock fronts generate clumps of high density. This density
enhancement can be large enough for the fluctuations to become
gravitationally unstable and collapse. This happens when the local
Jeans length becomes smaller than the size of the fluctuation.
However, the fluctuations in turbulent velocity fields are highly
transient.  The random flow that creates local density enhancements
can disperse them again.  For local collapse to actually result in the
formation of stars, Jeans-unstable shock-generated density
fluctuations must collapse to sufficiently high densities on time
scales shorter than the typical time interval between two successive
shock passages.  Only then are they able to `decouple' from the
ambient flow and survive subsequent shock interactions.  The shorter
the time between shock passages, the less likely these fluctuations
are to survive. Hence, the timescale and efficiency of protostellar
core formation depend strongly on the wavelength and strength of the
driving source (Klessen et al.\ 2000, Heitsch et al.\ 2001), and
accretion histories of individual protostars are strongly time varying
(Klessen 2001, hereafter paper III).

The velocity field of long-wavelength turbulence is found to be
dominated by large-scale shocks which are very efficient in sweeping
up molecular cloud material, thus creating massive coherent
structures. When a coherent region reaches the critical density for
gravitational collapse its mass typically exceeds the local Jeans
limit by far.  Inside the shock compressed region, the velocity
dispersion is much smaller than in the ambient turbulent flow and the
situation is similar to localized tur\-bulent decay.  Quickly a
cluster of protostellar cores forms. Therefore, models 1 to 3 with
zero support, decaying and large-scale turbulence, respectively, lead
to a {\em clustered} mode of star formation. The efficiency of
turbulent fragmentation is reduced if the driving wavelength
decreases. When energy is inserted mainly on small spatial scales, the
network of interacting shocks is very tightly knit, and protostellar
cores form independently of each other at random locations throughout
the cloud and at random times.  Individual shock generated clumps have
lower mass and the time interval between two shock passages through
the same point in space is small.  Hence, collapsing cores are easily
destroyed again and star formation is inefficient. This scenario
corresponds to the {\em isolated} mode of star formation.

This is visualized in figure~\ref{fig:3Dplot}, showing the density
structure of all five models at $t=0$ and at a time when the first
protostellar cores have formed by turbulent fragmentation and have
accreted roughly 30\% of the total mass.  For model 1 (Gaussian),
without turbulent support, the figure indicates at $t=0$ the initial
density distribution. For the other models, $t=0$ corresponds to the
phase of fully developed turbulence just before gravity is `switched
on' (in model 2 the driving mechanism is `switched off' at the same
time).  Time is measured in units of the global free-fall timescale
$\tau_{\rm ff} = (3\pi/32G)^{-1/2}\,\langle\rho\rangle^{-1/2}$, with
$\langle\rho\rangle$ being the mean density and $G$ the gravitational
constant. Dark dots indicate the location of dense collapsed core.  In
the non-supported model 1 all spatial modes are unstable initially and
in model 2 of decaying turbulence they become unstable after roughly
one crossing time. Therefore, these systems evolve into a filamentary
structure and protostellar cores form predominantly at the
intersections of the filaments.  Similarly, also large-scale
turbulence builds up a network of filaments, however, this time the
large coherent structures are not caused by gravity, but instead are
due to shock compression.  Once gravity is included, it quickly
dominates the evolution inside the shock compressed regions. The
random velocity component is quickly damped by dissipation, and again
a cluster of protostellar cores builds up. In the case of
intermediate-wavelength turbulence, cores form in small aggregates,
and small-scale turbulence predominantly results in the formation of
isolated cores. Note the different times needed for  mass
to be accumulated in dense cores.

\section{Mass Spectra from Turbulent Fragmentation}
\label{sec:mass-spectra}
\subsection{The Mass Spectra}
\label{subsec:mass-spectra}
Mass is the dominant parameter that determines stellar evolution. It
is therefore important to investigate the relation between the mass of
molecular clumps and protostellar cores, and the stars resulting from
the collapse of the former.  For the five models considered here,
figure \ref{fig:massspectra} plots the mass distribution of gas clumps
(thin lines), of the subset of Jeans-critical clumps (thin lines,
hetched distribution), and of collapsed cores (thick lines, hetched
area). It depicts four different evolutionary phases, the initial
distribution just when gravity is `switched on' (at $t=0$, left
column), and then after (turbulent) fragmentation has led to
protostellar core formation, i.e.\ when the fraction $M_{\rm *}$ of
the total mass accumulated in dense cores has reached values of
$M_{\rm *}\approx 5$\%, $M_{\rm *}\approx 30$\%, and $M_{\rm *}\approx
60$\% (columns 2 to 4, respectively). The clump mass spectra are
obtained applying a clump-finding algorithm similar to the one
described by Williams et al.\ (1994), but working on all three spatial
coordinates and adapted to make use of the SPH kernel smoothing
procedure (for details see Appendix 1 in paper I). To guide your eye,
two dotted lines indicate a slope $\nu = -1.5$ typical for the
observed power-law clump mass spectrum $dN/dM = M^{\nu}$, as well as
the Salpeter (1955) approximation $\nu = -2.33$ to the stellar IMF
appropriate for intermediate and high masses (e.g.\ Scalo 1998, Kroupa 2001).

\subsection{Time Evolution of Clump Mass Spectra in the Interplay
between Turbulence and Self-Gravity} 
\label{subsec:clump-mass-spectra}
In the initial phase, i.e.\ before gravity is `turned on' and local
collapse begins to set in, the clump mass spectrum (thin line) is not
well described by a single power law. The distribution has small width
and falls off steeply at larger masses. Below masses $M \approx 0.3
\,\langle M_{\rm J} \rangle$ the distribution becomes shallow, and
strongly declines at and beyond the resolution limit (indicated by a
vertical line). Clumps are on average considerably smaller than the
mean Jeans mass in the system $\langle M_{\rm J}\rangle$.  For masses
$M > 0.1\,\langle M_{\rm J} \rangle$ and for models 1 to 4, this
behavior resembles the spectrum of pre-stellar condensations found in
$\rho\,$-Ophiuchus (Motte et al.\ 1998, Johnstone et al.\ 2000, see also
Testi \& Sargent 1998 for Serpens).
Recall that for densities of $n({\rm H}_2) = 10^5\,$cm$^{-3}$ and
temperatures $T = 11.4\,$K, the mean Jeans mass in the system is
$\langle M_{\rm J} \rangle = 1\,$M$_{\odot}$.

In the later evolution the effects of gravitational attraction modify
the distribution of clump masses. Clumps merge and grow bigger, and
the mass spectrum extends towards larger masses.  At the same time the
number of cores which exceed the Jeans limit increases. Local collapse
sets in and results in the formation of dense cores. This happens
fastest and is most evident in model 1 which lacks turbulent
support. The velocity field is entirely determined by gravitational
contraction on all scales and at all times. The clump mass spectrum is
very well fit by a single power law and exhibits a slope $\nu \approx
-1.5$ as long as protostellar cores are forming and the overall
gravitational potential is dominated by non-accreted gas.

The influence of gravity on the clump mass distribution is weaker
where turbulence dominates over gravitational contraction on the
global scales, i.e.\ in model 2 during the early stages and in models
3 to 5 during all phases.  The more the turbulent energy dominates
over gravity, the more the spectrum resembles the initial case of pure
hydrodynamic turbulence. This is most extreme in model 5 of
small-wavelength turbulence, where the short interval between shock
passages prohibits efficient clump merging and the build up of a large
number of massive clumps. Only few clumps exceed the Jeans limit,
become gravitationally unstable, and collapse to form cores. The bulk of
the mass distribution remains unchanged by gravity and is never well
fit by a single power law. The mass spectrum retains the initial shape
with only few collapsed clumps added at the high-mass end.

\begin{figure*}[htp]
\unitlength1cm
\begin{picture}(18.0,16.0)
\put( 0.5, -1.5){\epsfxsize=17cm \epsfbox{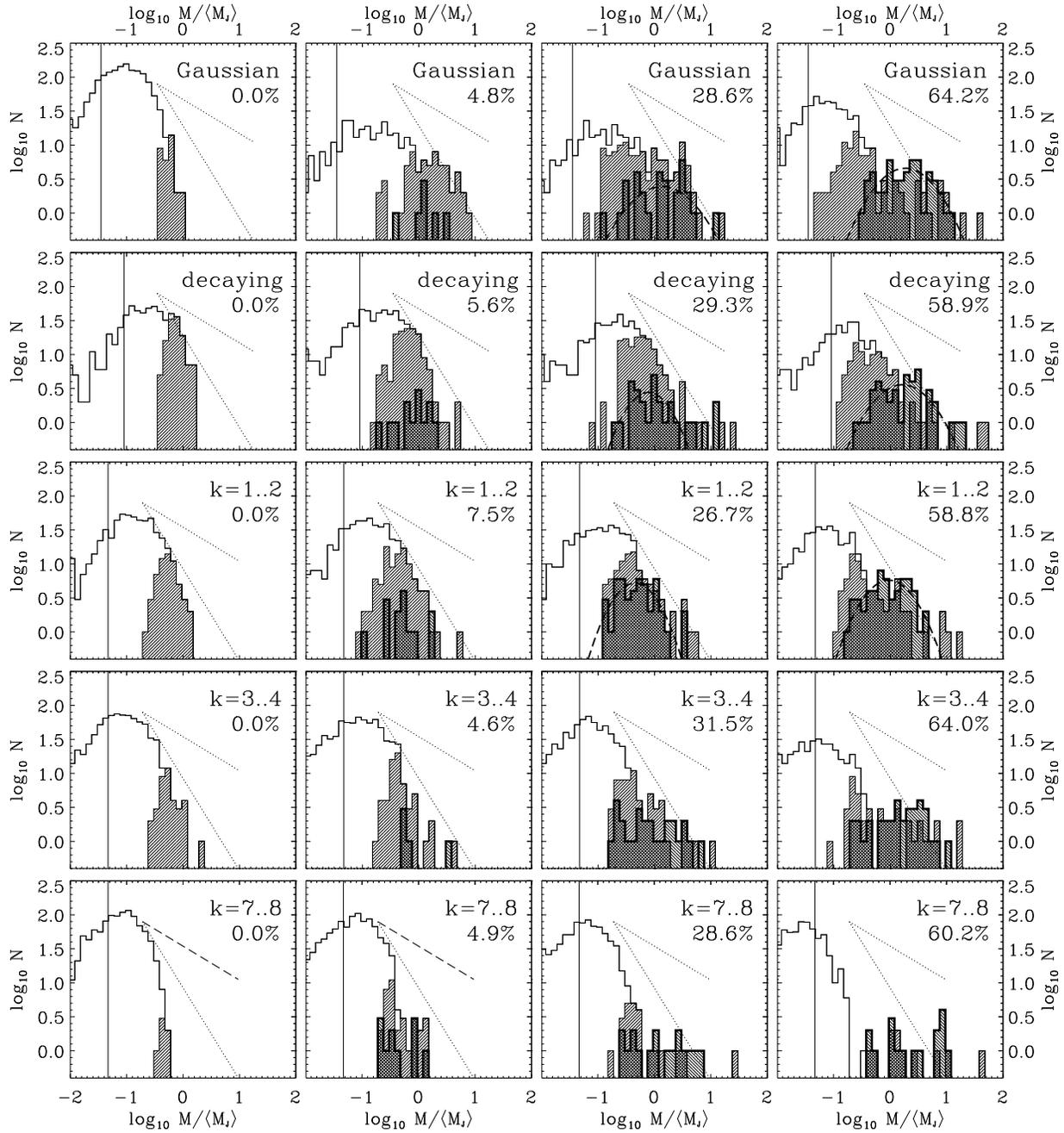}}
\end{picture}
\caption{\label{fig:massspectra} Mass
    spectra of gas clumps (thin lines), and of the subset of Jeans
    unstable clumps (thin lines, hatched distribution), and of dense
    collapsed cores (hatched thick-lined histograms).  Masses are
    binned logarithmically and normalized to the average thermal Jeans
    mass $\langle M_{\rm J}\rangle$. The left column gives the initial
    state of the system, just when gravity is `switched on', the
    second column shows the mass spectra when $M_{\rm *} \approx 5$\%
    of the mass is accreted onto dense cores, the third column
    describes $M_{\rm *} \approx 30$\%, and the last one $M_{\rm *}
    \approx 30$\%. For comparison with power-law spectra ($dN/dM
    \propto M^{\nu}$), a slope $\nu = -1.5$ typical for the observed
    clump mass distribution, and the Salpeter slope $\nu=-2.33$ for
    the IMF at intermediate and large masses, are indicated by the
    dotted lines in each plot. Note that with the adopted logarithmic
    mass binning these slopes appear shallower by $+1$ in the plot.
    The vertical line shows the SPH resolution limit. In columns 3 and
    4, the long dashed curve shows the best log-normal fit to the core
    mass spectrum. To compare the distribution of core masses with the
    stellar IMF, an efficiency factor of roughly $1/3$ to $1/2$ for
    the conversion of protostellar core material into single stars
    needs to be taken into account, as discussed in the text. For
    $T=11.4\,$K and $n({\rm H}_2) = 10^5\,$cm$^{-3}$, the average
    Jeans mass in the system is $\langle M_{\rm J}\rangle=
    1\,$M$_{\odot}$. Note, however, that the considered models can be
    scaled to different $\langle M_{\rm J}\rangle$ as well.}
\end{figure*}

When the scalelength of the dominant turbulent mode is increased, the
density structure becomes more coherent and clump mergers are more
frequent. The number of high-mass Jeans-unstable clumps increases,
yielding a wider clump mass distribution which exhibits a power-law
behavior for all masses larger than the resolution limit. In models 2
to 4, the slope lies in the interval $-2.5 < \nu \lesssim -2$.  For
individual models in figure \ref{fig:massspectra}, the slope $\nu$
increases with time as the statistical properties of the system become
more and more influenced by clump merging and gravitational
contraction onto high-density cores. When comparing similar
evolutionary phases for different models, again the clump spectrum
falls off less steeply if gravity dominates the evolution over larger
spatial scales, i.e.\ $\nu$ decreases from model 1 to model 5. When
$M_{\rm *} \approx 60$\% $\nu \approx -1.5$ for model 1, $\nu \lesssim
-2$ for models 2 and 3, while for models 4 and 5 a power-law
description is no longer sensible.  In summary, {\em the clump mass
  spectrum gets shallower when gravity becomes more important.} This
could explain the observed range of slopes $-1.9 \lesssim \nu \lesssim
-1.3$ for the clumps mass spectra in different molecular cloud regions
(e.g.\ Stutzki \& G{\"u}sten 1990, Williams et al.\ 1994, Heithausen
et al.\ 1998, Kramer et al.\ 1998, Onishi et al.\ 1998). The
importance of self-gravity for shaping the velocity and density
structure may differ from cloud to cloud. In the case of strong
gravity, their statistical properties furthermore depend sensitively
on the viewing angle (e.g.\ Klessen 2000).

In the late evolutionary stages, similar behavior holds for the subset
of Jeans-unstable clumps. For the three models of driven turbulence,
the distribution of the gravitationally supercritical clumps (as
indicated by hatched thin lines) is largest and widest for model 3 and
decreases in width and size towards model 5. The clump spectra of
models 2 and 3 are similar, indicating that the conditions for local
collapse within the coherent shock-compressed regions which result
from large-scale driving turbulence are comparable to turbulent
decay. Within these coherent structures, turbulence is strongly
reduced and stars formation is efficient. In the early stages of the
evolution, however, the mass spectrum of Jeans-unstable clumps is not
well described by a power law for {\em all} models, instead it is more
compatible with a log-normal distribution. For the Gaussian model the
peak is roughly at the average thermal Jeans mass $\langle M_{\rm
J}\rangle$ and decreases towards smaller masses when including
turbulence and decreasing the driving wavelength. The distribution of
Jeans-unstable clumps in model 5 peaks roughly at $1/4\,\langle M_{\rm
J}\rangle$. Thus small-scale turbulence produces clumps of on average
smaller mass scale than does large-scale turbulence.

\subsection{Protostellar Mass Spectra from Turbulent Fragmentation}
\label{subsec:core-mass-spectra}
Like the distribution of Jeans-unstable clumps, also the mass spectrum
of dense protostellar cores (thick hatched line) resembles a
log-normal in the model without turbulent support, and in the ones
with turbulent decay or long-wavelength turbulent driving. A
log-normal fit is obtained at times $M_{\rm *}\approx 30$\% and
$M_{\rm *}\approx 60$\%, and indicated by long-dashed lines in column
3 and 4. The corresponding mean values and widths are given in table
\ref{tab:log-normal}.  As for the Jeans-critical cores, the peak is
roughly at the {\em average thermal Jeans mass} $\langle M_{\rm
J}\rangle$ of the system. The width of the distribution spans two
orders of magnitude for the cluster size considered here, and is
approximately the same for all five models.  However, a log-normal fit
is only appropriate for models 1 to 3. The core mass spectrum for
models 4 and 5 is too flat, and a fit is not attempted. Here the
accretion histories of individual protostellar cores are not well
correlated, i.e.\ gas clumps typically do not contain multiple
protostellar cores. In this isolated mode of star formation, mutual
interaction and competition for gas accretion are not important.

A log-normal shape of the mass distribution may be explained by
invoking the central limit theorem (e.g.\ Larson, 1973, Zinnecker
1984, also Adams \& Fatuzzo 1996), as protostellar cores form and
evolve through a sequence of highly stochastic events, resulting from
the statistical nature of supersonic turbulence, stochastic clump
merging, and/or competitive accretion within merged clumps containing
multiple cores. The fact that the quality of the log-normal
description becomes bad for the case of isolated star formation (model
5) indicates that clump merging and competitive accretion are dominant
factors leading to a log-normal mass spectrum.  Isolated star
formation in supersonic turbulence exhibits a featureless flat mass
spectrum. It appears that the collective effects of gravitational
collapse and the synchronization of individual accretion histories
(i.e.\ the mutual interaction of protostellar cores and their
competition for accretion from a common gas reservoir) are necessary
to obtain mass spectra with features similar to the IMF
(\S\ref{subsec:comparison-IMF}). The fact that the distribution peaks
roughly at the average Jeans mass, despite strong variations of the
local Jeans mass for individual clumps which span a wide range of
densities, indicates that the system retains in a statistical sense
knowledge of its mean properties, even in the case of supersonic
turbulence.  The same behavior is seen in paper's I and II for a wide
varyity of Gaussian initial conditions. It is the average thermal
Jeans mass that introduces a scale into the mass function (e.g.\
Larson 1998, Elmegreen 1999).

\begin{deluxetable}{lccccc}
\tablehead{
\colhead{} &
\colhead{model 1} &
\colhead{model 2} &
\colhead{model 3} &
\colhead{model 4} &
\colhead{model 5} \\
\colhead{Parameter} &
\colhead{$\sim30$\%\phs$\sim60$\%} &
\colhead{$\sim30$\%\phs$\sim60$\%} &
\colhead{$\sim30$\%\phs$\sim60$\%} &
\colhead{$\sim30$\%\phs$\sim60$\%} &
\colhead{$\sim30$\%\phs$\sim60$\%} 
}
\tablecaption{\label{tab:log-normal}
Properties of Log-Normal Fit}
\startdata
$\log_{10} \mu$~~(peak)  & \phs{0.13}\phs\phs{0.26}&
{$-0.10$}\phs\phs{0.21}  & {$-0.34$}\phs{$-0.03$}&
\phs{---}\phs\phs\phs{---}& \phs{---}\phs\phs\phs\phs{---} \\
%
$\log_{10} \sigma$~~(width) & \phs{$0.52$}\phs\phs{$0.47$}&
\phs{$0.36$}\phs\phs{0.47}  & \phs{$0.36$}\phs\phs{$0.40$}&
\phs{---}\phs\phs\phs{---}& \phs{---}\phs\phs\phs\phs{---} \\
%
\enddata
\tablecomments{ The log-normal fits are obtained for models 1 to 3
only for two different evolutionary stages, when $M_{\rm *} \approx
30$\% and $M_{\rm *} \approx 60$\%. The applied functional form is
$dN/d\log_{10}M \propto
\exp\left[-0.5(\log_{10}M-\log_{10}\mu)^2/(\log_{10}\sigma)^2\right]$,
with mass $M$, mean $\mu$ and width $\sigma$ scaled to the average
Jeans mass in the system $\langle M_{\rm J} \rangle$. No fit has been
obtained for models 4 and 5, the core mass spectrum is too flat and
featureless.  When scaling the current models to physical units, the
width of the core mass distributon lies between the IMF estimates by
Miller \& Scalo (1979) and by Scalo (1986) and Kroupa et al.\ (1990),
who derive $\log_{10} \sigma = 0.67$ (their estimate with constant
star formation rate over $12\times 10^9$ years) and $\log_{10} \sigma
= 0.38$ (their model MS), respectively.  Recall that $\langle M_{\rm
J} \rangle = 1\,$M$_{\odot}$ for $n({\rm H}_2) =
10^5\,$cm$^{-3}$ and $T = 11.4\,$K. As the number of protostars in the
simulated cluster is limited to 50 -- 100, the comparison with the
stellar IMF applies to low to intermediate-mass stars only. The
statistics is not good enough for an investigation of the very
low-mass and the high-mass end of the IMF where the log-normal
parametrization fails.  }
\end{deluxetable}

\subsection{Comparison with the Stellar Mass Function}
\label{subsec:comparison-IMF}
For the direct comparison between the core mass spectrum and the IMF
one needs to adopt a physical scaling for the numerical model
(\S\ref{sec:numerics}). This is done here by taking the average
thermal Jeans mass to be $\langle M_{\rm J} \rangle =
1\,$M$_{\odot}$. Furthermore, one needs to estimate which fraction of
a typical protostellar core in the model will accrete onto the star(s)
in its interior, and which fraction may be expelled during the main
accretion phase by protostellar outflows, or be removed by tidal
effects in a clustered environment. For isolated cores forming single
stars, the mass loss due to radiation or outflows is expected to be
small and most of the core material will indeed end up in stars
(Wuchterl \& Tscharnuter 2000).  In this case the core mass spectrum
can be compared directly to the single-star IMF. If cores contain a
large amount of angular momentum, they are likely to form binary
stars. Indeed a initial binary fraction of almost 100\% is consistent
with observations of star clusters (Kroupa 1995). Assuming a more or
less uniform distributon of mass ratios (Duquennoy \& Mayor 1991)
leads to a shift of a factor of two in the characteristic mass of the
distribution compared to the single-star IMF. If the number of triple
systems is high, this shift could be even larger. To some degree this
effect can be taken into account by comparing the core mass spectrum
with IMF estimates that do not contain binary corrections (see Kroupa
2001 for a further discussion). In addition, a fraction of the
accreting matter may settle into a protobinary disk and may not
accrete onto the stars due to angular momentum conservation (e.g.\
Bate 2000).  In a cluster environment, this disk may be truncated and
leak out matter due to tidal interactions (e.g.\ Clarke, Bonnell, \&
Hillenbrand 2000). These effects depend strongly on the stellar
density of the cluster and its dynamical evolution. Altogether, the
peak in the mass spectrum of protostellar cores is expected to exceed
the characteristic mass in the single star IMF by a factor of 2 to 3
(i.e.\ is shifted by $0.3 \lesssim \Delta \log_{\rm 10}M \lesssim
0.5$). 

Additional uncertainty stems from the possible formation of O or B
stars in the stellar cluster. These would trigger the complete gas
removal due to ionizing radiation, therefore limiting the core
formation efficiency $M_{\rm *}$.  As it is not known in advance
whether at all and when high mass stars form during the protostellar
cluster evolution, the current models should be considered with
caution at very late phases $M_{\rm *}> 70$\%.

If one assumes a close correspondence between core masses and stellar
masses, then for models 1 to 3 the core mass distribution compares
well with the observed IMF at low to intermediate masses. The width of
the distribution falls right in between the log-normal IMF estimates
by Miller \& Scalo (1979) on one side and by Kroupa et al.\ (1990) and
Scalo (1986) on the other. The same holds when using the more common
multiple power-law description of the IMF (Scalo 1998 or Kroupa 2001,
see e.g.\ his figure 14). Also the characteristic masses in the
distribution become comparable if one adopts values of ${\rm few}
\times 0.1$ for the accretion efficiency from cores to the central
stars taking into account unresolved binaries and possible tidal
truncation of protobinary disks in a dense cluster environment as
mentioned before. However, more work needs to be done to obtain better
estimates for this efficiency factor. This requires combining the
simulations described here with detailed dynamical models of pre-main
sequence evolution, and extending the
current scheme to resolving larger density contrasts and larger
clusters sizes to be able to study in more detail the influence of
cluster environment on the dynamical evolution of accretion disks.

The protostellar clusters discussed here contain between 50 and 100
cores. This allows for comparison with the IMF only around the
characteristic mass scale, i.e.\ at low to intermediate masses. The
numbers are too small to study the very low- and high-mass end of the
distribution.  The same holds for the mass spectra discussed in paper
I and II.  For high-mass stars ($M \gtrsim 5\,$M$_{\odot}$) the
log-normal and the multiple power-law descriptions of the IMF begin to
differ significantly. The log-normal models predict too few high-mass
stars, and the same may also be true at the very low-mass end of the
IMF, in the brown dwarf regime (especially when taking binary
corrections into account).  Because of insufficient statistics at the
extreme ends of the distribution, the current set of simulations
cannot be used to {\em distinguish} between log-normal and power-law
IMF models. The log-normal fit at low to intermediate mass in figure 2
is therefore mainly attempted for the sake of simplicity, because only
two parameters, the peak value and the width, are sufficient to
characterize the distribution. Both can be conveniently compared with
the corresponding fit parameters for the stellar mass function in the
considered mass range.

The comparison reveals a striking contrast between the models of
turbulent decay or large-scale driving on the one side, and models of
short- to intermediate-wavelength turbulence on the other.  Models 1
to 3 lead to protostellar mass spectra that agree well with the
observations, whereas the mass spectra derived from models 4 and 5
compare only very poorly with the stellar IMF. They are too flat or
equivalently too wide. As small- to intermediate-scale turbulence
describe an isolated mode of star formation, this finding is
consistent with the hypothesis that most stars form in aggregates or
clusters (e.g.\ Adams \& Myers 2001).  To further constrain the
numerical models discussed here, it will be necessary to compute the
dynamical evolution of star-forming regions with 1000 protostellar
cores or more. Besides the width and characteristic mass of the core
distribution, also the detailed slope at very low and very high masses
and the apparent symmetry around the peak can then be included into
the analysis.

Finally, it needs to be noted that the current findings raise doubts
about attempts to explain the stellar IMF from the turbulence-induced
clump mass spectrum {\em only} (e.g.\ Elmegreen 1993, Padoan 1995,
Padoan et al.\ 2001, Padoan \& Nordlund 2001). Quite typically for
star forming turbulence, the collapse timescale of shock-compressed
gas clumps often is comparable to their lifetime (molecular cloud
clumps appear to be very transient, e.g.\ Bergin et al.\ 1997). This
not only has important consequences for the overall star formation
efficiency in turbulent clouds (Klessen et al.\ 2000), but more so for
the collapse behavior of individual Jeans-unstable shock-generated gas
clumps. While collapsing to form or feed protostars, clumps may loose
or gain matter from interaction with the ambient turbulent flow. In a
dense cluster environment, collapsing clumps may merge to form larger
clumps containing multiple protostellar cores, which subsequently
compete with each other for accretion form the common gas environment
(Bonnell 1997, paper's I and II). The resulting distribution of clump
masses in star forming regions strongly evolves in time (figure 2). In
dense clusters, furthermore, close encounters between accreting
protostars may become important leading to the expulsion of protostars
from the gas rich environment (as illustrated in figure 11 of paper
I). This terminates mass growth and if occuring frequently enough
modifies the resulting IMF. The mass accretion rates onto individual
protostars are highly stochastic and strongly depend on the cluster
environment (paper III).  For all this reasons, it is not possible to
infer a {\em one-to-one} relation between the clump masses resulting
from turbulent molecular cloud fragmentation and the stellar
IMF. Given our limited understanding of interstellar turbulence and
protostellar mass growth processes in dense clusters, the current
investigation (which attempts to include some of the above processes)
leads to the conclusion that -- although tempting in some cases -- it
is not appropriate to take a snapshot of the turbulent clump mass
spectrum as describing the IMF.

\section{Summary}
Stars form from turbulent fragmentation of molecular cloud material.
It is the relation between turbulent fragmentation, (localized)
gravitational collapse and star formation which is the focus of this
paper. As mass is the most important stellar parameter, particular
interest lies in the mass spectra of gas clumps, of the subset of
gravitationally unstable gas clumps, and of protostellar cores, the
latter one being the direct progenitors of stars. For this purpose
five numerical models of the evolution of self-gravitating isothermal
molecular gas have been analyzed, spanning the parameter range
relevant for molecular cloud dynamics.  In model 1 turbulent support
is not included and gravity is the dominant force shaping the velocity
and density structure. In model 2, initially supersonic compressible
turbulence is allowed to decay freely, and in models 3 to 5 supersonic
turbulence is continuously replenished on large, intermediate and
small scales, respectively, such that gravitational attraction is
compensated on global scales. In these models gravity is considered
only after turbulent equilibrium is established.

It has been shown in a previous study (papers I and II) that molecular
cloud regions without turbulent support form dense clusters of stars,
regardless of the initial density structure, within roughly one global
free-fall timescale. When gravitational contraction has sufficient
time to act, the clump mass spectrum is well approximated by a power
law $dN/dM\propto M^{\nu}$. The mass distribution of protostellar
cores, however, is better described by a log-normal with properties
similar to the observed IMF of multiple stellar systems for low and
intermediate-mass stars.

This analysis is extended here by more realistically considering
molecular cloud regions where turbulence is allowed to decay and where
it is continuously driven. Decaying turbulence leads to clustered star
formation much like in the case of pure gravitational
contraction. Supersonic turbulence, even if it is strong enough to
compensate gravity on large scales, will provoke {\em local} collapse
in shock compressed regions. As efficiency and timescale of star
formation depend sensitively on the strength and the spatial scale of
energy input into the system, large-scale turbulence leads to
clustered star formation on short timescales, whereas for small-scale
turbulence stars form in isolation and with low efficiency.

This is reflected in the overall clump mass spectrum. The clump mass
spectrum of pure hydrodynamic turbulence is not well described by a
single power law, its width is too small compared to the observed data
and it is too steep at the high-mass end. This changes, when gravity
is taken into account. Clumps merging and accumulation of matter
through local collapse, lead to a clump mass spectrum that extends to
larger masses and that exhibits power-law behavior. The more strongly
the overall dynamical evolution is influenced by gravity and
synchonized, coherent collapse behavior, the flatter the power
spectrum becomes. In the extreme case of pure gravitational
contraction the clump mass distribution exhibits a slope $\nu \approx
-1.5$. For the case of turbulence decay and large-scale injected
turbulence the slope is $\nu \lesssim -2$ during the intermediate
phases of the dynamical evolution. In the case of small-scale
turbulence, the infuence of gravity is weak and the clump mass
distribution remains steep, close to the spectrum of purely
hydrodynamic turbulence (i.e.\ before gravity was `switched on' in the
model). The dependence of the slope of the clump mass spectrum on the
relative importance of gravity may explain the range of observed
power-law indices in different molecular clouds regions, as one
expects the ratio between self-gravity and turbulent kinetic energy to
vary from cloud to cloud.

Molecular cloud properties that result in clustered star formation
lead a stellar mass spectrum that is well fit by a log-normal at low
and intermediate masses. The distribution exhibits a maximum close to
the average Jeans mass in the system. This indicates that the system
somehow retains `knowledge' of its mean properties, even in the case
of supersonic compressible turbulence.  The mean thermal Jeans mass in
a cloud indeed introduces a characteristic mass scale to clustered
star formation.  For regions where turbulence is decaying or driven on
large scales only, it appears that the collective effects of
gravitational collapse and the correlation between individual
accretion histories are necessary to obtain mass spectra with features
similar to the IMF.  Isolated star formation, on the contrary, as
implied by turbulence that is driven on small scales, yields a
featureless flat spectrum.  This is in agreement with the hypothesis,
that most stars form in aggregates and clusters.

Shock-generated clumps in interstellar turbulence are highly
transient. Their average lifetime in the turbulent flow is of the same
order of their collapse timescale. In a dense cluster environment,
furthermore competitive accretion and mutual protostellar interactions
are important effects. The current investigation shows that it is
therefore not possible to infer a one-to-one relation between
turbulent clump mass spectra and the stellar IMF.

\acknowledgements I thank Peter Bodenheimer, Andreas Burkert, Fabian
Heitsch, Pavel Kroupa, Doug Lin, Mordecai-Mark Mac~Low, and
G{\"u}nther Wuchterl for many stimulating discussions on star
formation and the IMF and/or fruitful collaboration. I furthermore
appreciate the comments and suggestions of the referee John
Scalo. They helped to clarify the arguments presented here.  I
acknowledge financial support by a Otto-Hahn-Stipendium from the
Max-Planck-Gesellschaft and partial support through a NASA
astrophysics theory program at the joint Center for Star Formation
Studies at NASA-Ames Research Center, UC Berkeley, and UC Santa Cruz.


\end{document}